\documentclass[12pt]{iopart}

\usepackage{bm}
\usepackage{cite}
\usepackage{graphicx}
\usepackage{amsfonts}
\usepackage{iopams}
\usepackage{setstack}

\begin{document}

\title{Two-spin entanglement distribution near factorized states}

\author{Fabrizio Baroni$^1$, Andrea Fubini$^{1,2}$, Valerio Tognetti$^{1,3}$, Paola Verrucchi$^{1,4}$}
\address{$^1$ {Dipartimento di Fisica dell'Universit\`a di
Firenze, Via G. Sansone 1, I-50019 Sesto F.no (FI), Italy}}
\address{$^2$ {CNISM,
UdR Firenze, Via G. Sansone 1, I-50019 Sesto F.no (FI), Italy}}
\address{$^3$ {Istituto Nazionale di Fisica Nucleare Sezione
di Firenze, Via G. Sansone 1, I-50019 Sesto F.no (FI), Italy}}
\address{$^4$ {Centro di Ricerca e Sviluppo "Statistical Mechanics and 
Complexity" dell'INFM-CNR}, sezione di Firenze, Via G. Sansone 1, I-50019 
Sesto F.no (FI), Italy}

\date{\today}
\begin{abstract}

We study the two-spin entanglement distribution along the infinite $S=1/2$ 
chain described by the $XY$ model in a transverse field; closed analytical 
expressions are derived for the one-tangle and the concurrences $C_r$, $r$ 
being the distance between the two possibly entangled spins, for values of 
the Hamiltonian parameters close to those corresponding to factorized 
ground states. The total amount of entanglement, the fraction of such 
entanglement which is stored in pairwise entanglement, and the way such 
fraction distributes along the chain is discussed, with attention focused 
on the dependence on the anisotropy of the exchange interaction.
Near factorization a characteristic length-scale naturally emerges in the system, 
which is specifically related with entanglement 
properties and diverges at the critical point of the fully isotropic 
model.
In general, we find that anisotropy rule a complex 
behavior of the entanglement properties, which results in the fact that 
more isotropic models, despite being characterized by a larger amount of 
total entanglement, present a smaller fraction of pairwise entanglement: 
the latter, in turn, is more evenly distributed along the chain, to the extent
that, in the fully isotropic model at the critical field, the 
concurrences do not depend on $r$. 
\end{abstract}

\maketitle

\section{Introduction}
The analysis of entanglement properties has recently furnished new
insights into several peculiar features of many-body systems, such as
the occurrence of quantum phase transitions, or that of non trivial
factorized ground
states\cite{ArnesenBV01,Osterlohetal02,OsborneN02,Vidaletal03,Guetal03,
Verstraeteetal04,JVidaletal04,Roscildeetal04,DusuelV05,Roscildeetal05,Camposetal06}.
Different types of
entanglement can be defined in many-body systems, but computable
measures are available just for a few of them. In this sense, a
privileged role is played by the bipartite entanglement of
formation\cite{Bennettetal96}, and by the related quantities,
one-tangle and concurrence, which represent the entanglement of formation 
between one qubit and the rest of
the system, and that between two selected qubits of the system,
respectively. In particular the definition of the concurrence holds
not only for pure states, as in the case of the one-tangle, but also for 
mixed ones~\cite{Hilletal97,Wootters98}.

When magnetic systems are considered, the qubit is naturally represented 
by a spin with $S=1/2$: for interacting magnetic models described by 
Hamiltonians 
with certain symmetry properties, both the one-tangle and the concurrence 
are expressed in terms of standard magnetic observables, such as the 
magnetizations and the correlation functions, making it feasible a 
quantitative analysis of the entanglement dependence on the Hamiltonian 
parameters. If analytical expressions are available, a general discussion 
of such dependence is at hand, which is the reason why a renewed interest 
is being devoted to exactly solvable models.

In this paper we focus on the $T=0$ behavior of the best known 
one-dimensional XY model in a transverse field, in the vicinity of 
factorized ground states\cite{Kurmannetal82,Roscildeetal05}. We 
derive closed analytical formulas for the magnetization and the 
correlators, as functions of the anisotropy, the field, and the distance 
between the two selected spins, which allow us to study the long-distance 
behavior of the concurrence as the Hamiltonian parameters are varied.
By using these expressions for the correlation functions, we explicitly proof 
the divergence of the range of the concurrence in the anisotropic model (whose
preliminary result was presented in Ref.\cite{Amicoetal06}) and we extend it in
the whole parameter space, studying the fully isotropic case as well as the
slightly anisotropic region close to the spin saturation.
Our analysis, besides analytically confirming the divergence of the range 
of the concurrence, shows that to such 
divergence corresponds the appearance of a characteristic length-scale in 
the system, that we have named {\it two-spin entanglement length}. This 
length-scale depends on the value of the anisotropy and keeps finite as 
far as the model belongs to the Ising universality class, while diverging 
when factorization gets to coincide with saturation, i.e. for the 
isotropic $XX$ model, which belongs to the Kosterlitz-Thouless 
universality class.
Correspondingly, the way the two-spin entanglement distributes
along the chain is found to strongly depend on the symmetry of the model:
in the XX model a good amount of entanglement can be stored even between 
two spins which are far apart from each other, while in the Ising model the 
pairwise entanglement of the ground state is shared only between nearest 
and next-nearest neighboring spins.

The structure of the paper is as follows: In Sec.~\ref{s.Model} we 
introduce the model and the entanglement properties we aim at studying; In 
Sec.~\ref{s.LongdC} we study the long-distance pairwise entanglement both 
in the anisotropic and in the isotropic case; In Sec.~\ref{s.EntLength}
we define and analyze the two-spin entanglement length, while in 
Sec.~\ref{s.ResRatio} we use our results to understand the interplay 
between pairwise entanglement and multipartite entanglement. Finally, in 
Sec.~\ref{s.Conclusions}, we draw the conclusions.

\section{Model}
\label{s.Model}
The XY model in a transverse field is described by the Hamiltonian
\begin{equation}
\label{e.XY}
H=J\sum_{i}\left[(1+\gamma)S^x_i S^x_{i+1}+(1-\gamma)S^y_i
S^y_{i+1}-h S^z_i\right]\,,
\end{equation}
where $i$ runs over the sites of an infinite chain, $S_i^{\eta}$
($\eta{=}x,y,z$) are the $\,S=1/2$ quantum spin operators,
$\gamma{\in}[0,1]$ is the anisotropy, and $h=g\mu_BH/J$ is the reduced
magnetic field; $J>0\,$ is the strength of the exchange interaction.

For $0<\gamma\le 1$ the model belongs to the Ising universality class and 
at $T=0$ the critical field $h_c=1$ separates a disordered phase 
($h>h_c$), from a 
spontaneously broken-symmetry phase, where the staggered order parameter 
is finite ($\langle S^x_i\rangle\neq0$). In the isotropic case, 
$\gamma=0$, the model has an addictional rotational symmetry on the $xy$ 
plane, and the critical field coincides with the saturation field, above 
which all the spins incoherently align parallel to the field.  For 
$\gamma=0$ and $h\ge 1$ the system is in a fully-polarized phase ($\langle 
S^z_i\rangle=\frac{1}{2}$), while for $h<h_{\rm c}$, the systems is in a 
gapless phase with $\langle S^z_i\rangle<\frac{1}{2}$, $\langle 
S^x_i\rangle=0$ and power-law decaying correlation functions in the $xy$ 
plane. No spontaneous symmetry breaking is present in the isotropic case, 
as testified by $\langle S_i^x \rangle$ being null for whatever value of 
the applied field.

Let us now consider the $h-\gamma$ parameter space of the Hamiltonian 
(\ref{e.XY}): The ground state of the model is exactly 
factorized~\cite{Kurmannetal82} 
\begin{equation}
|GS\rangle=\prod_{i}|\phi_i\rangle~,
\label{e.factGS}
\end{equation}
on the circle $h^2+\gamma^2=1$, as well as along 
the line $\{h\ge 1, \gamma=0\}$: Such ground state has a N\`eel 
structure given by 
$|\phi_i\rangle=(-1)^i\cos\theta_\gamma|\uparrow_i\rangle+
\sin\theta_\gamma|\downarrow_i\rangle~$,
with $\cos\theta_\gamma=\sqrt{(1-\gamma)/(1+\gamma)}\equiv\alpha$,
which reduces to the trivial ferromagnetic ground state for $\gamma=0$
and $h\ge1$.
In what follows, we will refer to the circle $h^2+\gamma^2=1$ as the {\it 
factorized circle}, and to the line $\{\gamma=0$, $h\ge 1\}$ as the {\it
factorized line}.

For the model Eq.~(\ref{e.XY})
the concurrence $C_r$ between two spins sitting on sites
$i$ and $j$, with  $|i-j|=r$, reads~\cite{Amicoetal04}
\begin{eqnarray}
C_r&=2\max{\{0,C_r',C_r''\}},\label{e.C}
\\
C_r'&=|g_r^{xx}+g_r^{yy}|-\sqrt{(\frac{1}{4}+g_r^{zz})^2-M_z^2},
\label{e.C'}
\\
C_r''&=|g_r^{xx}-g_r^{yy}|+g_r^{zz}-\frac{1}{4} \label{e.C''}~,
\end{eqnarray}
while the one-tangle is $\tau_1 =1-4(M_x^2+M_z^2)~$,
with the correlators $g_{ij}^{\eta\eta}=\langle S_i^\eta
S_j^\eta\rangle$ and the magnetizations $M_\eta=\langle
S^\eta_i\rangle$~.  The terms $C_r'$ and
$C_r''$, Eqs.~(\ref{e.C'}-\ref{e.C''}) are related to the probabilities 
for the two considered
spins to be either in antiparallel or in parallel Bell states,
respectively\cite{Fubinietal06}. The total amount of bipartite
entanglement may be estimated by the so called {\em 
two-tangle}, namely the sum $\tau_2=2\sum_{r}C_r^2~$,
which is related with the one-tangle via the monogamy inequality $\tau_2\le
\tau_1$\cite{CoffmanKW00,OsborneV05}. 
The difference $\tau_1 -\tau_2$ is the so called {\it residual tangle}, 
while the ratio $\tau_2/\tau_1$ is usually referred to as the 
{\it entanglement ratio}.

Eqs.~(\ref{e.C}-\ref{e.C''}) has been originally derived~\cite{Amicoetal04} exploiting the symmetries of the Hamiltonian Eq.~(\ref{e.XY}), for this reason in presence of spontaneous symmetry breaking, i.e. for $\gamma>0$ and $h<1$ one has to be careful. This problem has been studied in Refs.~\cite{Syljuasen03,Osterlohetal06}: Eqs.~(\ref{e.C}-\ref{e.C''}) generally hold when $C_r''>C'_r$ - i.e. for $h^2+\gamma^2>1$~\cite{Amicoetal06} - while in the antiparallel region
$h^2+\gamma^2<1$ where $C'_r>C''_r$ they represent a lower bond for the pairwise
entanglement. Moreover, from Eq.~(7) of Ref.~\cite{Osterlohetal06}, one can see that in the asymptotic limit $r\to \infty$ they stay valid also for 
$C'_r>C''_r$, which makes the analysis of the long-distance 
concurrence, reported below, valid both inside and outside the factorized 
circle. As for the XX model, no spontaneous symmetry breaking 
occurs for whatever value of the field.


\section{Long distance concurrence}
\label{s.LongdC}
\subsection {Anisotropic case}
\label{ss.anisotropic}

Let us first consider the behavior of the model for $\gamma{>}0$ and 
$h>h_{\rm f}$, in the vicinity of the factorized circle: we will keep 
fixed and finite the value of $\gamma$, and vary the field, meaning that 
we will move along horizontal lines in the $h-\gamma$ plane.

The $T=0$ correlation functions entering the expressions of $C_r$ 
for the $XY$ model in a transverse field are usually evaluated
numerically, by computing the corresponding Toeplitz determinants, and
cannot be written in closed form for generic $r$, except in the case
of factorized ground states, where they do not depend on $r$. Since we
are interested in the behavior of the concurrence as the factorized circle 
is approached, we fix the value of $\gamma$ and derive
$C_r$ as a series expansion in the difference
$h-h_{\rm f}$, with the factorizing field $h_{\rm
f}=\sqrt{1-\gamma^2}$. 
The entries of Toeplitz determinant are basically given by the well known $G$ function~\cite{BarouchMD71}
\begin{equation}
G(r,h,\gamma)=
\frac{1}{\pi}\int_0^{\pi}
d\phi
\frac{(h{-}\cos\phi)\cos(r\phi){+}\gamma\sin\phi\sin(r\phi)}{\lambda(h,\gamma;\phi)}~,
\label{e.G}
\end{equation}
with $\lambda(h,\gamma;\phi)=\sqrt{(h-\cos\phi)^2+\gamma^2\sin^2\phi}$, and 
we have to expand it in the difference $h-h_{\rm f}$, 
thus obtaining 
\begin{eqnarray}
G(r,h,\gamma)&=&\frac{1}{\pi}\int_0^{\pi}\!\!\!d\phi
\frac{\left({\sqrt{1{-}{\gamma }^2}}{-}\cos\phi\right)\cos(r\phi){+}
\gamma\sin\phi\sin(r\phi)}{\lambda_{\rm f}(\phi)}{+}\nonumber
\\
&~&~~{+}\frac{1}{\pi}\int_0^{\pi}\!\!\!d\phi
\Bigg[\frac{\cos(r\phi)}{\lambda_{\rm f}(\phi)}\left(1{-}
\frac{({\sqrt{1{-}{\gamma}^2}}{-}\cos\phi)^2}
{\lambda_{\rm f}^2(\phi)}\right){-}\\
&~&~~~~~~~~~~{-}\frac{\left({\sqrt{1{-}{\gamma 
}^2}}{-}\cos\phi\right)
\gamma\sin\phi\sin(r\phi)}{\lambda_{\rm f}^3(\phi)}
\Bigg](h{-}h_f)+\cdots~,\nonumber
\end{eqnarray}
with $\lambda_{\rm f}(\phi)=\lambda(h_{\rm f},\gamma;\phi)$. 
The trigonometric functions of the angle $r\phi$ can be represented in terms of powers of these functions, then, integrating by part and resumming all the $r$ terms, one gets\cite{Baronitesi06}
\begin{eqnarray}
G(r,h,\gamma)&=
\frac{1}{2\gamma}\alpha^r(h-h_{\rm f})+O(h-h_{\rm f})^2~,\nonumber\\
G(0,h,\gamma)&=
\alpha+\frac{1}{2\gamma}(h-h_{\rm f})+O(h-h_{\rm f})^2~,
\label{e.Gexp}\\
G(-r,h,\gamma)&=
-\frac{2\gamma}{1{+}\gamma}\alpha^{-r-1}{+}
\frac{1{+}2r\gamma{-}(2r^2+1)\gamma^2}{2\gamma(\gamma{+}1)^2}\alpha^{r-2}(h{-}h_{\rm f})
{+}O(h{-}h_{\rm f})^2~,\nonumber
\end{eqnarray}
where the first and the third equations hold for $r\neq0$.
From Eqs.~(\ref{e.Gexp}), the expansion for the correlators are found in 
closed form:
\begin{eqnarray}
g_r^{xx}&=\frac{(-1)^r}{4}\left[\frac{2\gamma}{1+\gamma}+
\frac{\alpha^{2r+1}-2\alpha}{2\gamma}(h-h_{\rm f})\right]+
O(h-h_{\rm f})^2~,
\label{e.gxx}
\\
g_r^{yy}&=-\frac{(-1)^r}{4}\frac{\alpha^{2r-1}}{2\gamma}(h-h_{\rm f})
+O(h-h_{\rm f})^2~,
\label{e.gyy}
\\
g_r^{zz}&=\frac{1}{4}\left[\alpha^2+
\left(\frac{\alpha}{\gamma}+
\frac{\alpha^{2r-1}}{\gamma+1}\right)(h-h_{\rm f})\right]+
O(h-h_{\rm f})^2~,
\label{e.gzz}
\end{eqnarray}
and the magnetization along the field direction reads
\begin{equation}
M_z=\frac{\alpha}{2}+\frac{1}{4\gamma}(h-h_{\rm f})+O(h-h_{\rm 
f})^2\,.
\label{e.mz}
\end{equation}
In the most anisotropic $\gamma=1$ case, it is $\alpha=0$ and the only
finite correlator up to the first order in $(h-h_{\rm f})$ is
$g^{xx}_r$, whose modulus gets the maximum value
(i.e. $|g^{xx}_r|=1/4$) independently of $h$ and $r$; 
the first correction is of order $(h-h_{\rm f})^2$,
being $|g^{xx}_r|=1/4 -(h-h_{\rm f})^2/16$.  
Notice that
Eqs.~(\ref{e.gxx}-\ref{e.mz}) do not hold for $\gamma=0$, where in
fact they display unphysical singularities.

Similar expressions are found for $h<h_{\rm f}$ and, by defining the {\it 
distance} $\varepsilon\equiv|h-h_{\rm f}|$, we obtain the first-order term 
of the expansion in $\varepsilon$ for $C_r$, which reads
\begin{equation}
C_r=\frac{\alpha^{2r-1}}{2\gamma}\varepsilon+O(\varepsilon^2)~,
\label{e.Crfirst}
\end{equation}
and noticeably holds for whatever $r$.

Let us now focus on the 
range $R$ of the concurrence\cite{Fubinietal06,Amicoetal06}, 
which is the distance 
between the two farthest entangled spins along the chain, i.e.
\begin{equation}
R~~{\rm :}\quad C_r>0
~,~\forall r\le R \quad \wedge 
\quad C_r=0
~,~\forall r>R. 
\label{e.R}
\end{equation}
\begin{figure}
\begin{center}
\includegraphics[width=8.5cm]{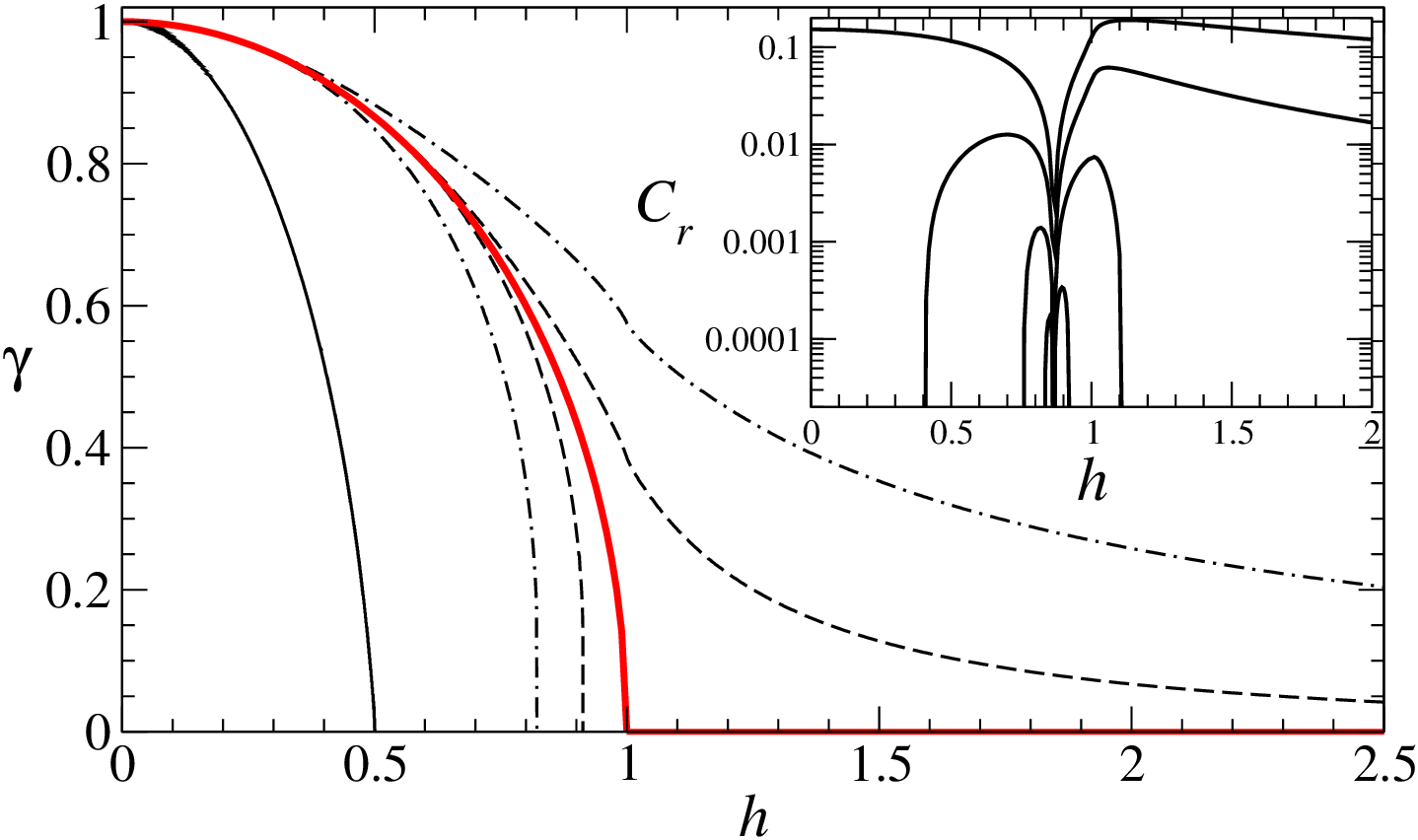}
\caption{Entanglement phase diagram: The thick curve is the line where the 
ground state is exactly factorized. The regions between curves with
the same drawing represent the areas of the $h-\gamma$ plane where
$C_r\neq 0$ for $r=3$ (dot-dashed) and $r=4$ (dashed). $C_2=0$ only below the solid line and at $h=h_{\rm f}$ and $C_1$ vanishes only at $h=h_{\rm f}$. The inset shows $C_r$ vs $h$ for different $r=1,...,4$ (from top to bottom) at
$\gamma=0.5$.} \label{f.fase}
\end{center}
\end{figure}
Since $C_r$ is finite for all $r$ at the first order in $\varepsilon$,
$R$ diverges for $\varepsilon\to 0$. This statement is compatible with
the behavior depicted by the exact numerical data shown in  
Fig.~\ref{f.fase}; from the same data, we also see that for whatever 
$r>1$, it exists a distance $\varepsilon_0(r,\gamma)$ such that $C_r>0$ 
for $0<\varepsilon<\varepsilon_0(r,\gamma)$. 
On the other hand, since Eq.~(\ref{e.Crfirst}) cannot describe the 
vanishing of $C_r$ at $\varepsilon_0(r,\gamma)$, in order to further 
analyze the 
behavior of $R$, one has to evaluate the concurrence up to the second 
order in $\varepsilon$. 
As we are interested in the behavior of the
long-distance concurrence, we are allowed to use the large-$r$
asymptotic expressions of the correlators~\cite{BarouchMD71}, thus
finding
\begin{equation}
\frac{C_r}{2}=\frac{\alpha^{2r-1}}{4\gamma}\varepsilon-\left[A^2-\delta
A^2(r)\right]\varepsilon^2+O(\varepsilon^3)~,
\label{e.Crsecond}
\end{equation}
where $A^2=\alpha^2 (\gamma +3)/32\gamma^3$, and $\delta
A^2(r)\sim O(r^{-2})$.  It is important to notice that, in contrast to 
Eq.~(\ref{e.Crfirst}), the above expression only
holds for large $r$. The behavior predicted by Eq.~(\ref{e.Crsecond}),
though approximated, is consistent with that shown by the exact
numerical data (see Fig.~\ref{f.Crsecond}). In particular, beyond the
trivial zero in $\varepsilon=0$, Eq.~(\ref{e.Crsecond}) has another
zero which approximates $\varepsilon_0(r,\gamma)$:
\begin{equation}
\varepsilon_0(r,\gamma)\simeq\frac{\alpha^{2r-1}}{4\gamma A^2}\,,
\label{e.epsilon0}
\end{equation}
where we neglected the $\delta A^2(r)$ term, which vanishes for
$r\to\infty$. Notice that the symmetry of Eqs.~(\ref{e.Crsecond}) 
and (\ref{e.epsilon0}) with 
respect to the sign of the difference $h-h_{\rm f}$ arises from 
the $2^{\rm nd}$ order expansion in $h-h_{\rm f}$ itself, that 
becomes more and more accurate for larger $r$ (see 
Fig.~\ref{f.Crsecond}).
\begin{figure}
\begin{center}
\includegraphics[width=7cm]{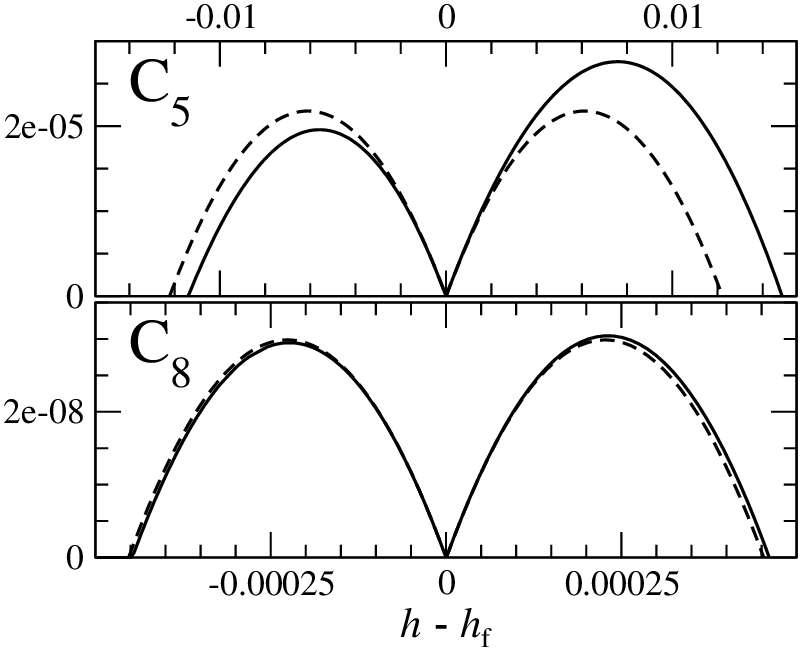}
\caption{$C_r$ versus $h-h_{\rm f}$, for $\gamma = 0.5$,
$r=5$ (top panel) and
$r=8$ (bottom panel): Comparison between the
exact (full line) and approximated [Eq.~(\ref{e.Crsecond}) with  $\delta
A^2(r)=0$] value
(dashed line).} \label{f.Crsecond}
\end{center}
\end{figure}

For a given (large) $r$, $\varepsilon_0(r,\gamma)$ is
the distance from $h_{\rm f}$ at which $C_r$ gets finite while
approaching the factorizing field.  We can rephrase this statement by
saying that, for fixed $h\neq h_{\rm f}$, the farthest entangled spins
are those whose distance $r$ fulfills Eq.~(\ref{e.epsilon0}), with
$\varepsilon_0=|h-h_{\rm f}|$.  Therefore, if we consider $r$ as a
continuous variable, Eq.~(\ref{e.epsilon0}) can be inverted and we
obtain, for the range of the concurrence defined in Eq.~(\ref{e.R}),
the following expression
\begin{equation}
R\simeq \frac{1}{\ln\alpha^2}
\ln\varepsilon+
\frac{\ln(4\alpha\gamma
A^2)}{\ln\alpha^2}\quad{\rm for}\quad
\varepsilon \to 0~.
\label{e.Rasymptotics}
\end{equation}

From the above expression we see that, for a fixed value of $\varepsilon$, 
a larger anisotropy implies a smaller value of $R$. On the other hand, 
when the 
anisotropy increases, one should pay particular attention to the overall 
consistency of the reasoning, as the existence itself of $\varepsilon_0$ 
is not generally due for small $r$, given that 
Eq.~(\ref{e.Rasymptotics}) holds only for large $r$. 
Specifically, 
for $h<h_{\rm f}$ $C_1$ is always finite, and for $h>h_{\rm f}$ both $C_1$ 
and $C_2$ keeps finite no matter the value of the field, as seen in 
Fig.~1~. In particular, for $\gamma=1$ the above scheme breaks down: For 
any finite value of the field, the only non-zero concurrences are those 
between nearest and next-nearest neighbors, as from the exact 
results by Pfeuty\cite{PfeutyP70}, which give 
\begin{equation*}
C_1= \frac{h^2}{8}+\frac{3h^4}{128} +O(h^6)~~;
~~~~ C_2=\frac{h^4}{128} +O(h^6)~.
\end{equation*}

\subsection {Isotropic case} 
\label{ss.isotropic}

Factorization and quantum criticality are two distinct phenomena, 
occurring usually for different values of the external magnetic field, 
being typically $0<h_{\rm f}<h_{\rm c}$. When factorization occurs, 
peculiar features of the two-spin entanglement distribution are observed 
at $h_{\rm f}$ where, in turn, standard magnetic observables behave quite 
trivially. On the other hand, it is just the peculiar behavior of these 
latter properties that signals the occurrence of a quantum 
phase transitions at $h_{\rm c}$, where two-spin entanglement distribution  
has no distinctive features.

When factorization and quantum criticality get to coincide at 
$h_{\rm f}=h_{\rm c}$, standard  two-point correlation 
functions and pairwise entanglement together signal 
the occurrence of a phenomenon which corresponds both to a factorization 
(though of a particular type, i.e. saturation) and to a quantum phase 
transition (though of topological type rather than second order).

The specificity of the above depicted situation translates into a 
peculiar distribution of the two-spin entanglement along the chain, which 
gives the isotropic $XX$ model a special role in the overall analysis, as shown 
below.

\begin{figure}[b]
\begin{center}
\includegraphics[width=8.5cm]{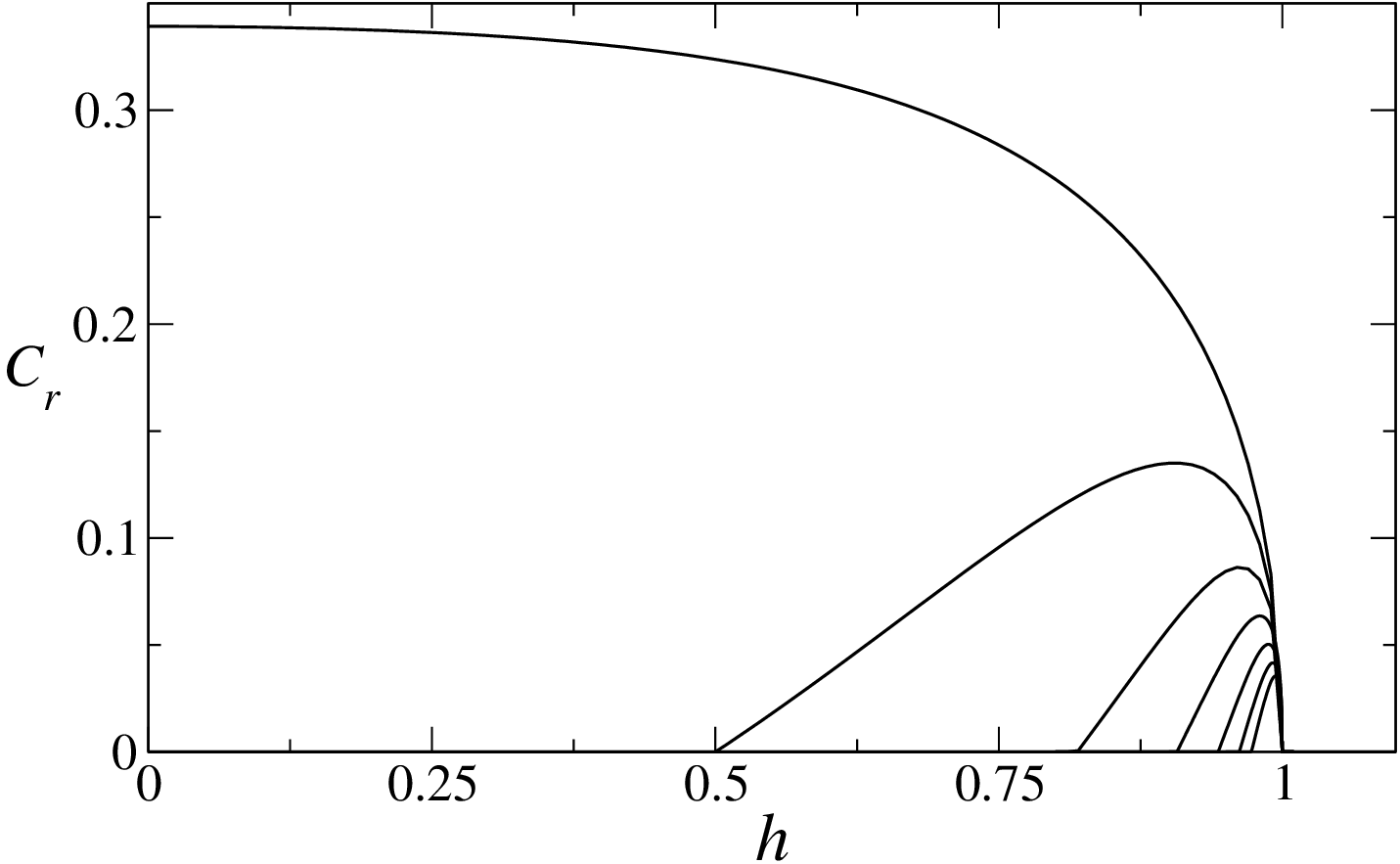}
\caption{concurrence $C_r$ vs $h$ for $\gamma=0$ ($XX$ model) and 
$r=1,...,7$ (from the highest to the lowest curve).} 
\label{f.gamma0}
\end{center}
\end{figure}

Let us consider Eq.~({\ref{e.XY}) with $\gamma =0$, in the 
non-trivial quasi-ordered phase, $h<h_{\rm f}=1$:
Using a procedure similar to that depicted above,
we obtain the following closed forms for the expansions of the 
correlators

\begin{eqnarray}
g_r^{xx(yy)}&={(-1)^r}
\bigg[
\frac{\varepsilon^{1/2}}{\pi\sqrt{2}}-
\frac{(4r^2-1)}{12\pi\sqrt{2}}\varepsilon^{3/2}+
\frac{2r(r^2-1)}{9\pi^2}\varepsilon^{2}
\bigg]
+O(\varepsilon^{5/2})~,
\label{e.XXgxx}
\\
g_r^{zz}&=\frac{1}{4}-\frac{\sqrt{2}}{\pi}\varepsilon^{1/2}-
\frac{\varepsilon^{3/2}}{6\pi\sqrt{2}}+
\frac{4 r^2}{3\pi^2}\varepsilon^{2}+O(\varepsilon^{5/2})~,
\label{e.XXgzz}
\end{eqnarray}
where $g_r^{xx}=g_r^{yy}$ due to the symmetry in the $xy$ plane, 
and
the magnetization along the field direction reads
\begin{equation}
M_z=\frac{1}{2}-\frac{1}{\pi}\cos^{-1} h~.
\label{e.XXmz}
\end{equation}
The expansion of the concurrence consequently reads
\begin{eqnarray}
C_r&=&\frac{2\sqrt{2}}{\pi}\,\varepsilon^{1/2}-
\frac{4r}{\pi\sqrt{3}}\,\varepsilon+
\frac{8r\sqrt{3}{-}(4r^2{-}1)\pi}{3\pi^2\sqrt{2}}\,\varepsilon^{3/2}+
\nonumber\\
&~&+2r\frac{30\sqrt{3}
+20(r^2{-}1)\pi+(4r^2{-}5)\sqrt{3}\pi^2)}{45\pi^3}\,\varepsilon^{2}+
O(\varepsilon^{5/2})~.
\label{e.XXCrexp}
\end{eqnarray}

From the above expressions, we see that the change of the universality 
class at $\gamma=0$ has drastic effect on the mechanism of rearrangement 
of two-spin entanglement along the chain. In particular, the fact 
that the correlation functions have an algebraic dependence on $r$, 
rather than the exponential one found in the anisotropic case, reflects 
in the independence of $r$ of the first term of Eq.~(\ref{e.XXCrexp}). 
Thus, whatever the selection of the two spins in the chain they 
share the same amount of entanglement.
Such invariance is quite a surprising feature and it suggests,
according to the analysis proposed in Ref.~\cite{Facchietal06}, 
the more relevant role of multipartite entanglement in the isotropic case.

Moreover, the comparison between Eqs.~(\ref{e.XXCrexp}) and 
(\ref{e.Crsecond}), as well as that between the inset of Fig.~\ref{f.fase} 
and Fig.~\ref{f.gamma0}, evidences an overall increase of {\it all} the 
concurrences $\{C_r\}$ in the most isotropic case, consistently with what 
is observed in the XXZ model\cite{Guetal03}. 

Let us now study $R$ as saturation is approached 
from below, i.e. for $h\to h_{\rm f}^-$: Its divergence is 
favored by the increase of the symmetry, as testified by the 
singularity
of the prefactor $1/\ln\alpha^2$ in Eq.~(\ref{e.Rasymptotics}) 
as $\gamma\to0$, that
signals a qualitative change in the behavior of $R$. In contrast to the
anisotropic case, for $\gamma=0$ we already got $C_r$ to order $\varepsilon^2$
[Eq.~(\ref{e.XXCrexp})] and we do not have to resume the $1/r$ asymptotic
expansions of the correlators to evaluate the range of the concurrence. The
farthest entangled spins are those whose distance $r$ fulfills
Eq.~(\ref{e.XXCrexp})=0; the latter is an equation of the 3rd order in $r$ that
for sufficiently small $\varepsilon$ has three real solutions, the smallest
positive one is just $R$. As in previous large $r$
studies based on the XXZ model~\cite{Amicoetal06,Fubinietal06}, we find that $R$ diverges more 
rapidly than in the anisotropic case [Eq.~(\ref{e.Rasymptotics})], namely
\begin{equation}
R \propto \varepsilon^{-1/2}~.
\label{e.XXR}
\end{equation}

For $h>1$, being the ground state factorized, all
the entanglement measures vanish; however one may fix a value of $h$
larger than unity, and study the behavior of the concurrences as
$\gamma\to 0$. We therefore computed the correlators and the
concurrences in this parameter region as a function of the
anisotropy. In Fig.~\ref{h1.5.gammaRangeC}, 
$\{C_r\}_{r=1,...,6}$ are plotted versus
$\gamma$ at fixed field. Again we see that, while
approaching the factorized ground state, i.e. for $\gamma \to 0$, all
the $\{C_r\}$ get finite and the range of the concurrence diverges
with a logarithmic trend: $R \propto 1/\ln\gamma$. Our results show that
this behavior is general for any $h>1$ and the divergence becomes more
and more pronounced for $h\to 1$.
\begin{figure}
\begin{center}
\includegraphics[width=8.5cm]{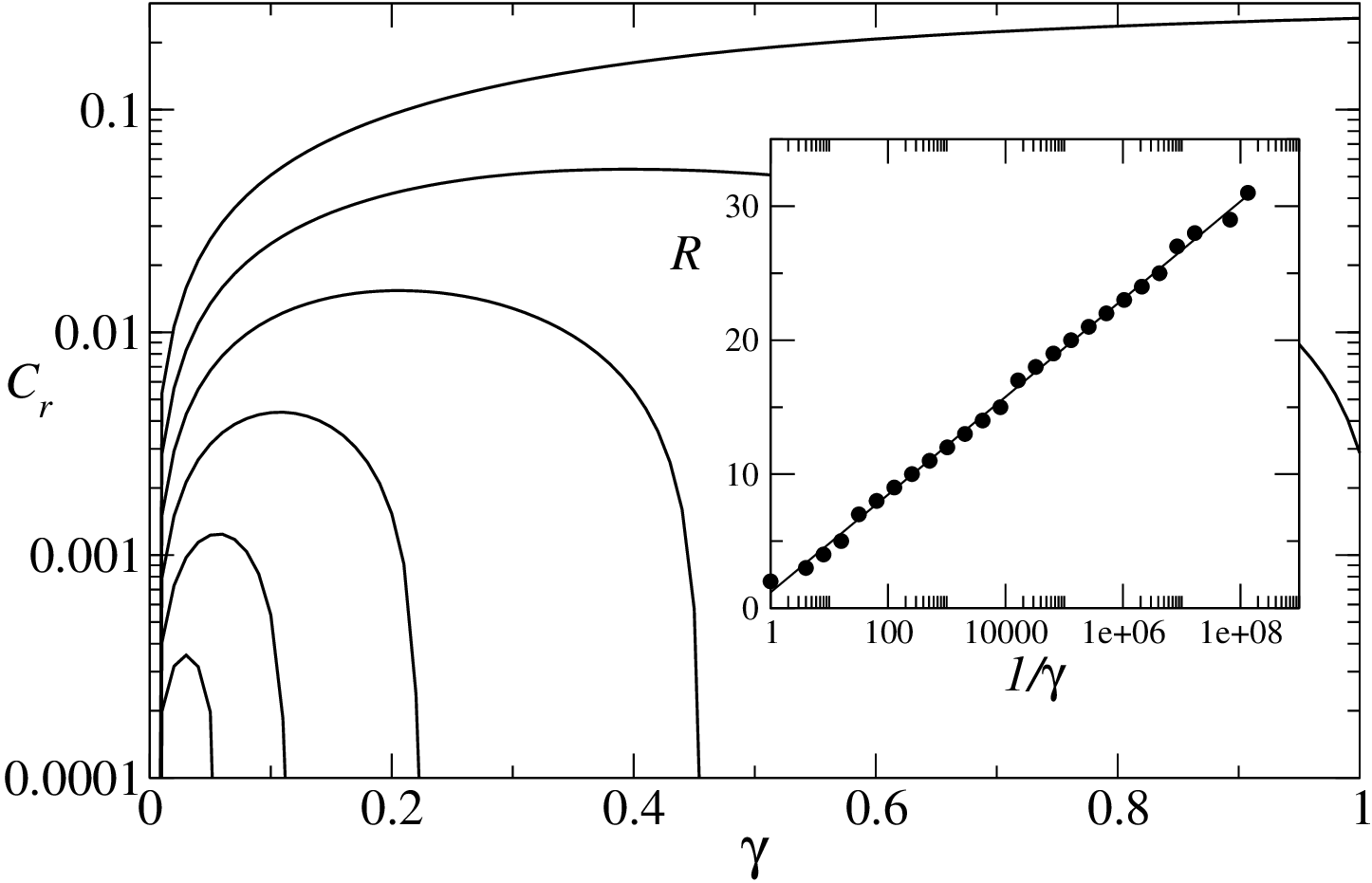}
\caption{$C_r$ versus $\gamma$ for $h=1.2$ and $r=1,...,6$ (from the highest 
to the lowest curve). The inset shows the logarithmic divergence of R
for  $\gamma \to 0$.} \label{h1.5.gammaRangeC}
\end{center}
\end{figure}
\begin{figure}[b]
\begin{center}
\includegraphics[width=8.5cm]{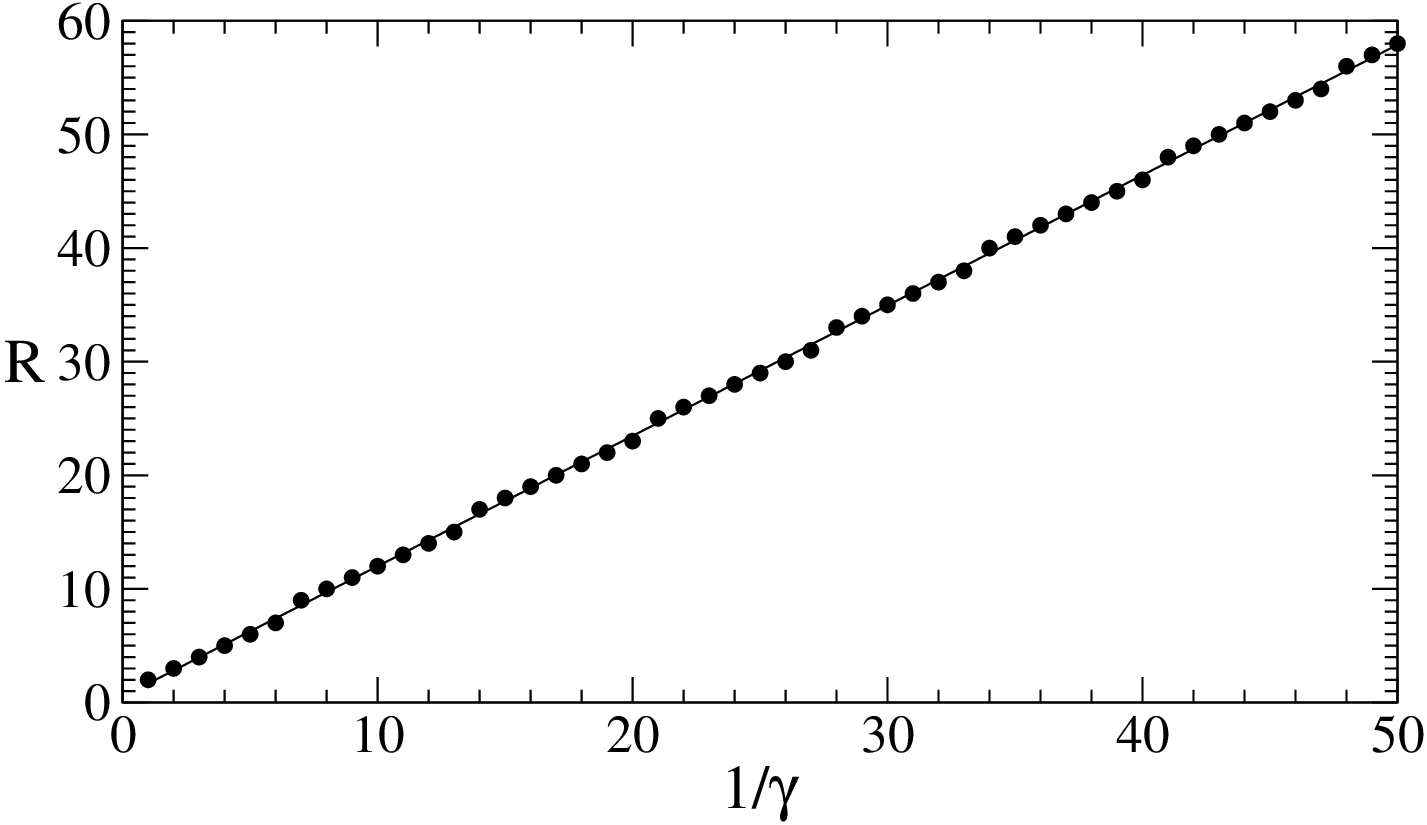}
\caption{$R$ versus $1/\gamma$ for $h=h_{\rm c}=1$.} \label{f.Rvsgamma}
\end{center}
\end{figure}
Eventually, as shown by Fig.~\ref{f.Rvsgamma}, at the saturation field 
$h_{\rm f}=h_{\rm c}=1$ the divergence of the concurrence range 
modifies its dependence
on the anisotropy, being $R\propto 1/\gamma$. Thus, the change of the
character of the divergence of R from logarithmic to 
power-law observed both for $\gamma=0$ and $h\to h_{\rm f}^-$ and 
for $h=h_{\rm f}$ and $\gamma\to 0$ suggests the critical point of the 
isotropic model to represents a peculiar point in this context, as also 
proposed in 
Ref.~\cite{Franchinietal07}, though in a slightly different sense.

\section{Two-spin entanglement length $\xi_{_{\rm 2SE}}$}
\label{s.EntLength}

The most noticeable feature of the first-order expansion in
$\varepsilon$ of $C_r$, Eq.~(\ref{e.Crfirst}), is the purely
exponential dependence on $r$, $C_r\sim \alpha^{2r}$, which indicates 
that a characteristic length emerges in the system near factorized ground 
states. This length scale is
\begin{equation}
\xi_{_{\rm 2SE}}\equiv -\frac{r}{\ln(\alpha\gamma 
C_r)}=\frac{1}{|\ln\alpha^2|}~,
\label{e.xi2se}
\end{equation}
and we have named it {\it two-spin entanglement length} as it specifically 
characterizes the distribution of entanglement between different spin 
pairs along the chain. In fact, by looking at the expansions of the 
correlators, Eqs.~(\ref{e.gxx}-\ref{e.gzz}), we notice that 
$g^{yy}_r$ has the same purely exponential behavior of $C_r$, which means 
that $\xi_{_{\rm 2SE}}$ coincides with the standard correlation length 
along the less favored direction which, in turn, does not enter the 
characterization of the magnetic behavior. In fact, the relation between 
$g^{yy}_r$ and $C_r$ is not accidental: as a matter of fact the change of 
sign of $g^{yy}_r$, namely $g^{yy}_r<0$ for $h<h_{\rm f}$, $g^{yy}_r>0$ for $h>h_{\rm f}$, and $g^{yy}_r=0$ at $h_{\rm f}$, rules the swap between 
parallel and antiparallel entanglement, i.e. between $C_r'$ and $C_r''$ [see Eqs.~(\ref{e.C'}) and (\ref{e.C''})].
So that surprisingly $g^{yy}_r$,
which is the less significant correlator as far as the 
standard magnetic properties are concerned, plays a relevant role in 
determining spin pair entanglement properties close to the factorization.

The above definition Eq.~(\ref{e.xi2se}) makes sense only if
Eq.~(\ref{e.Crfirst}) holds, i.e. for very small $\varepsilon$, where
the concurrence is finite for any spin pair along the chain. In other 
terms,  when $R$ diverges $C_r$ is found to decay exponentially with a 
characteristic length $\xi_{_{\rm 2SE}}$ which gets larger and larger as 
the anisotropy decreases, finally diverging as the isotropic critical 
point $(h=1,\gamma=0)$
is approached along a vertical line, being
\begin{equation}
\xi_{_{\rm 2SE}} \sim \frac{1}{2\gamma} \qquad {\rm for} \qquad
\gamma\to 0~.
\label{e.csigamma0}
\end{equation}

In order to further investigate the behavior of the isotropic model, 
let us extend our analysis to the vicinity of the factorized line:
In the previous Section we observed analogies between the behavior of
the two-spin entanglement close to the factorized circle (i.e. $h<1$)
and close to the factorized line (i.e. $h>1$). In particular we found
that also for $h>1$ and $\gamma\to 0$ all the concurrences $\{C_r\}$ become
finite, so that one can ask whether it is possible to extend the
definition of $\xi_{_{\rm 2SE}}$ to the region $h>1$ and $\gamma\ll 1$.
We computed $C_r$ versus $r$ for a fixed value of $\gamma\ll1$ and
also in this case we found an exponential dependence on $r$, as shown
by the inset of Fig.~\ref{f.xi2se}.
\begin{figure} 
\begin{center}
\includegraphics[width=8.5cm]{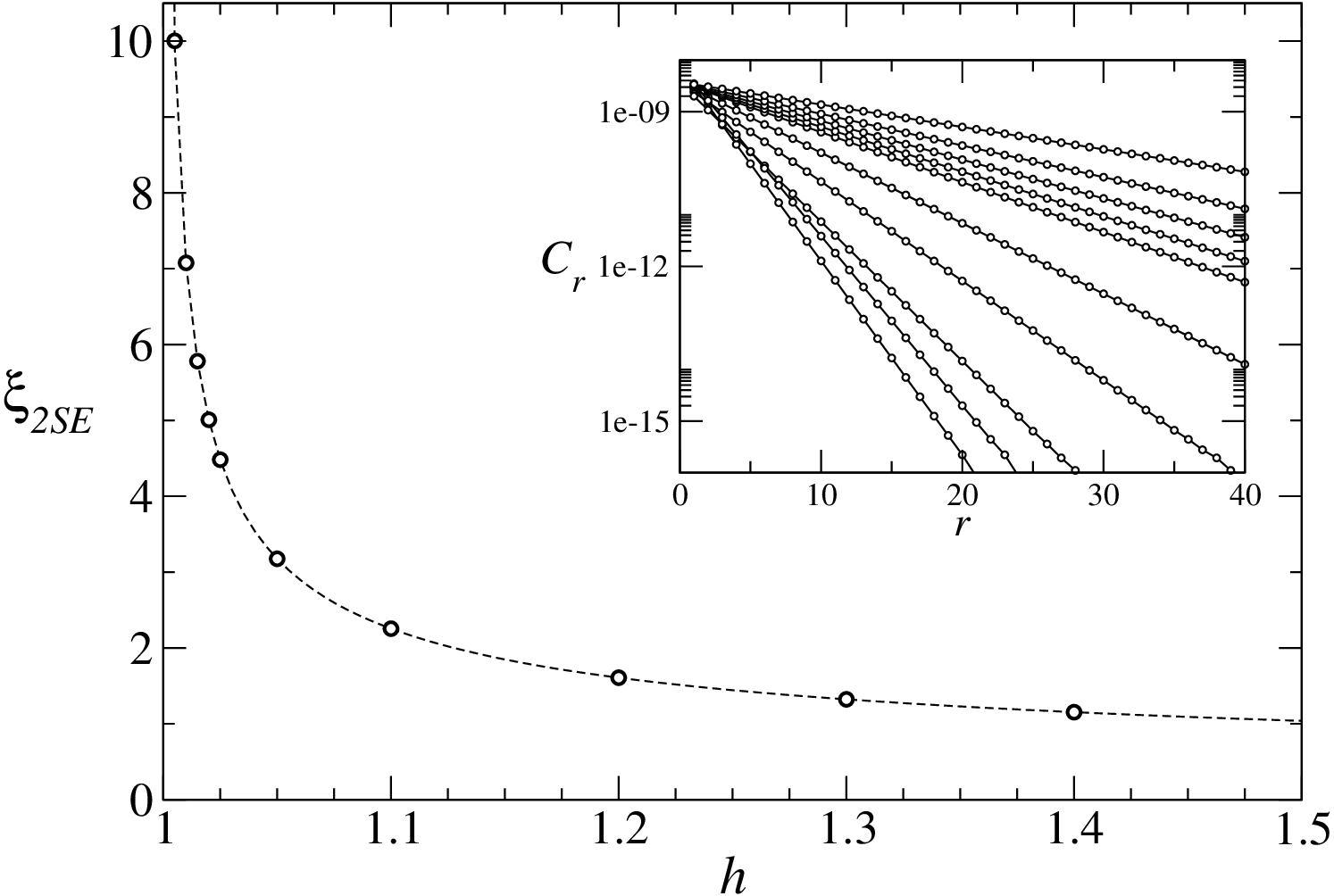}
\caption{$\xi_{_{\rm 2SE}}$ vs $h$ at $\gamma=7.5\,10^{-9}$. 
The dashed-line is the best fit $f(h)=0.055 + 0.69/(h-1)^{0.50}$. In
the inset the lin-log plot of $C_r$ vs $r$, the slopes of the lines
correspond to the values of $\xi_{_{\rm 2SE}}$ in the main panel.}
 \label{f.xi2se}
\end{center}
\end{figure}
Thus, extending the definition~(\ref{e.xi2se}) to the region $h>1$,
the two-spin entanglement length can be evaluated by the slope of the
lines in a lin-log plot of $C_r$ vs $r$. As the point $\gamma=0$ and
$h=1$ is approached from $h>1$, our numerical results show that
\begin{equation}
\xi_{_{\rm 2SE}} \propto \frac{1}{(h-1)^{\nu}} \qquad {\rm for} \qquad
h\to 1^+~,
\label{e.xiC_FL}
\end{equation}
with $\nu=0.50$. We notice that, by using the identity 
$\gamma=\sqrt{1-h^2_{\rm f}}$,  one may recast Eq.~(\ref{e.csigamma0}) in 
the form $\xi_{_{\rm 2SE}}\propto 1/[2(1-h^2_{\rm f})^{1/2}]\sim 
1/[2\sqrt{2}(1-h_{\rm f})^{1/2}]$ which reproduces the behavior of 
the above Eq.~(\ref{e.xiC_FL}): This tells us that the way the two-spin 
entanglement length diverges while approaching the ($\gamma=0,h=1$) 
critical point does not depend on whether one moves $h$ 
or $\gamma$.

The divergence of the two-spin entanglement length for $\gamma\to 0$
and $h\to1$ means that in the neighborhood of this point, not only all
the concurrences $\{C_r\}$ are finite for any $r$, but also that the
pairwise entanglement does not depend on the distance $r$ between
spins, as the expression for the concurrences Eq.~(\ref{e.XXCrexp}) 
anticipated. Elsewhere the concurrence is 
either vanishing for short distances or exponentially suppressed with $r$. 
The behavior of $\xi_{_{\rm 2SE}}$ confirms the peculiarity of the
critical point of the isotropic model among those where the ground state of
the system gets factorized. We understand such peculiarity as related with 
the fact that the critical point of the isotropic model does in fact 
coincides with saturation, i.e. with a special case of factorization.
Therefore, while peculiar features of entanglement properties and standard 
magnetic properties are usually observed  for different values of the 
field, $h_{\rm f}$ and $h_{\rm c}$ respectively, in the isotropic model 
they occur together at the saturation field. A significant consequence of 
this feature is that, due to the lack of anisotropy in 
the XY plane, the XX model has $g^{xx}_r=g^{yy}_r$ and the correlation 
lengths along the $x$ and $y$ direction are consequently identical: 
Therefore, the two spin entanglement length, that we have found to equal the 
correlation length along the $y$ direction, in the isotropic, $\gamma=0$, 
model coincides with the relevant correlation length along the $x$ 
direction, with which consequently shares the divergence at 
$h_{\rm c}=h_{\rm f}$.

\section{Residual entanglement and entanglement ratio}
\label{s.ResRatio}

The amount of entanglement stored between two spins far apart in the chain does
not only rely on the distribution of concurrences $\{C_r\}$, but also on the
total entanglement of the system and on whether it is bipartite or
multipartite. A simple way to investigate this issue is to evaluate the
one-tangle $\tau_1$, the residual tangle $\tau_1-\tau_2$, and the relative 
weight of
the pairwise entanglement through the entanglement ratio $\tau_2/\tau_1$, 
in the
neighborhood of the factorized circle. For $0<\gamma\le 1$, using the
definitions of one- and two-tangle and the expressions for the magnetizations
and concurrences,
one obtains
\begin{eqnarray}
\tau_1&=\frac{(1-\gamma)(3+\gamma)}{8\gamma^3(1+\gamma)}\varepsilon^2+
O(\varepsilon^3)~, 
\label{e.one}
\\
\tau_1-\tau_2&=\frac{(1-\gamma)^2(2+\gamma)}{8\gamma^3(1+\gamma)}\varepsilon^2+
O(\varepsilon^3)~,
\label{e.res}
\\
\frac{\tau_2}{\tau_1}&=\frac{(1+\gamma)^2}{3+\gamma}+O(\varepsilon)~.
\label{e.ratio}
\end{eqnarray}

For $\gamma=0$, as noticed above, the correlators do not decay 
exponentially with $r$, but they rather follow a power law, as a 
consequence of the quasi-long range order characterizing the ground state 
for $h<1$ [compare Eqs.~(\ref{e.gxx}-\ref{e.gzz}) with 
Eqs.~(\ref{e.XXgxx}) and (\ref{e.XXgzz})]. This behavior, which reflects 
on the concurrences Eq.~(\ref{e.XXCrexp}), makes it cumbersome to evaluate 
the sum in the two-tangle expression. In fact, it is not 
difficult to show that $\tau_2\propto\varepsilon^{1/2}$, but in order to 
obtain the proportionality coefficient with good accuracy one should 
retain several terms in the small $\varepsilon$ expansion of $C_r$. 
For this reason we preferred to numerically 
compute $\tau_2$ and $\tau_2/\tau_1$ close to the factorizing field up to 
$\varepsilon=10^{-6}$ and we verified that the first term of 
Eq.~(\ref{e.ratio}) holds also for $\gamma=0$.

From the analysis of the above expressions, we notice that the larger the
anisotropy the higher the relative weight of the pairwise entanglement close to
the factorized ground state. In particular, in the pure Ising limit $\gamma=1$
the entanglement in the ground state is totally stored in pairwise form 
up to order $\varepsilon^2$, but both the total entanglement and the 
residual tangle are strongly suppressed, 
being $\tau_1 = \varepsilon^4/32+O(\varepsilon^{6})$ and 
$\tau_1-\tau_2=\varepsilon^6/64$. 
In the opposite limit, small values of the anisotropy $\gamma$ favor the
presence of multipartite entanglement and reduce the relative weight of the
two-spin entanglement. In order to reconcile this last statement with the 
fact that, as shown in Section~\ref{ss.isotropic}, a smaller anisotropy implies 
larger $C_r$, one should notice that, as the anisotropy of
the model decreases, the one-tangle becomes larger and larger and, for 
$\gamma=0$ it is
\begin{eqnarray}
\tau_1&=1-4M_z^2 \nonumber\\
&=\frac{4\sqrt{2}}{\pi}\varepsilon^{1/2}-\frac{8}{\pi^2}\varepsilon+
O(\varepsilon^{3/2})~,
 \label{e.XXone}
\end{eqnarray}
where $M_z$ is given by Eq.~(\ref{e.XXmz}).
Thus, even if in the $\gamma\to0$ limit the fraction of pairwise 
entanglement reduces, the two-spin entanglement takes advantage of the 
overall increase of the total amount of entanglement stored in the ground 
state.

\section{Conclusions}
\label{s.Conclusions}

In this paper we studied how the two-spin entanglement is distributed 
along the spin chain described by the Hamiltonian (\ref{e.XY}), focusing 
our attention on the possible connections between the pairwise
entanglement spreading and the symmetry of the model. In particular we 
have considered the neighborhood of factorized ground
states, where all the concurrences 
get finite, no matter the distance between the two spins, and the range of 
the concurrence diverges 
[see Eqs.~(\ref{e.Rasymptotics}) and (\ref{e.XXR})].
We have derived closed analytical expressions for correlators and 
concurrences in the
neighborhood of the factorized circle: Using these formulas we have derived 
an analytical expressions for the range of the 
concurrence $R$, and shown that, whenever $R$ diverges, a characteristic 
length scale $\xi_{_{\rm 2SE}}$, Eq.~(\ref{e.xi2se}) naturally emerges in the 
system. This two-spin entanglement length, which defines the pure 
exponential decay of $C_r$, evidenced by Eq.(\ref{e.Crfirst}), is 
finite for finite anisotropy and diverges as the isotropic critical point 
is approached, where it in fact coincides with the magnetic correlation 
length on the plane perpendicular to the applied field.

%

Our description shows how the two-spin entanglement distribution along the 
chain evolves, while moving from the Ising ($\gamma=1$) to the isotropic
($\gamma=0$) model:  In fact, for $\gamma\to 1$,
it results that a finite amount of entanglement can be stored between 
nearest neighbor spins only; in addition in the pure Ising case the 
range of the concurrence is always finite.
On the other hand, the fully isotropic exchange interaction of the 
$\gamma=0$ model evidently favors the entanglement
rearrangement also between distant spins.

Remarkably enough, despite the concurrence is either vanishing for short
distances or exponentially suppressed with $r$ in any point of the parameter
plane $h-\gamma$, in the isotropic case, close to $h=h_{\rm c}=1$ 
all the concurrences $\{C_r\}$ are finite and their value is independent of the
specific pair of spins along the chain whatever the distance $r$ between them.
This fact together with the change of the divergence character of the
concurrence range testifies the special role played by the critical point of the
isotropic model in the distribution of the entanglement between two spins of the
system.

\section{Acknowledgments}
The authors would like to acknowledge L.~Amico, A.~Cuccoli, F.~Illuminati 
and S.~Pascazio for useful discussions and the PRIN2005029421 project 
for financial support.

\end{document}